# Observation of a phonon avalanche in highly photoexcited hybrid perovskite single crystals

G. M. Vanacore[1*§], J. Hu[2,3*§], E. Baldini[4*§], C. A. Rozzi[5], M. Amato[6], H. Wei[7], J. Huang[7], S. Polishchuk[8,9], M. Puppin[8,9], A. Crepaldi[9,10], M. Grioni[9,10], M. Chergui[8,9], F. Carbone[9,11], and A. H. Zewail[12†]

[1] Laboratory of Ultrafast Microscopy for Nanoscale Dynamics (LUMiNaD), Department of Materials Science, University of Milano-Bicocca, Via Cozzi 55, 20125 Milano (Italy).

[2] Laboratory for Shock Wave and Detonation Physics, Institute of Fluid Physics, China Academy of Engineering Physics, Mianyang, Sichuan 621900, China.

[3] State Key Laboratory for Environment-Friendly Energy Materials, Southwest University of Science and Technology, Mianyang, Sichuan 621010, China

[4] Department of Physics, Massachusetts Institute of Technology, Cambrigde (MA), USA

[5] CNR – Instituto Nanoscienze, Modena, Italy.

[6] Université Paris-Saclay, CNRS, Laboratoire de Physique des Solides, 91405 Orsay, France

[7] Department of Applied Physical Sciences, University of North Carolina-Chapel Hill, Chapel Hill (NC), USA

[8] Laboratory of Ultrafast Spectroscopy, ISIC, École Polytechnique Fédérale de Lausanne (EPFL), CH-1015 Lausanne, Switzerland

[9] Lausanne Centre for Ultrafast Science (LACUS), École Polytechnique Fédérale de Lausanne (EPFL), CH-1015 Lausanne, Switzerland

[10] Institute of Physics, École Polytechnique Fédérale de Lausanne (EPFL), CH-1015 Lausanne, Switzerland

[11] Laboratory for Ultrafast Microscopy and Electron Scattering (LUMES), Institute of Physics (IPHYS), École Polytechnique Fédérale de Lausanne (EPFL), Lausanne, Switzerland.

[12] Physical Biology Center for Ultrafast Science and Technology, Arthur Amos Noyes Laboratory of Chemical Physics, California Institute of Technology, Pasadena (CA), USA.

*Authors contributed equally to this work.
[§]Correspondence to: giovanni.vanacore@unimib.it, jianbo.hu@caep.cn, ebaldini@mit.edu

[†]Deceased August 2nd 2016.

## Abstract

In hybrid lead halide perovskites, the coupling between photogenerated charges and the ionic degrees of freedom plays a crucial role in defining the intrinsic limit of carrier mobility and lifetime. However, direct investigation of this fundamental interaction remains challenging because its relevant dynamics occur on ultrashort spatial and ultrafast temporal scales. Here, we unveil the coupled electron-lattice dynamics of a $CH_3NH_3PbI_3$ single crystal upon intense photoexcitation through a unique combination of ultrafast electron diffraction, time-resolved photoelectron spectroscopy, and time-dependent *ab initio* calculations. We observe the structural signature of a hot-phonon bottleneck effect that prevents rapid carrier relaxation, and we uncover a phonon avalanche mechanism responsible for breaking the bottleneck. The avalanche involves a collective emission of low-energy phonons – mainly associated with the organic sub-lattice – that proceeds in a regenerative manner and correlates with the accumulation and confinement of photocarriers at the crystal surface. Our results indicate that in hybrid perovskites carrier transport and spatial confinement are key to controlling the electron-phonon interaction and their rational engineering is relevant for future applications in optoelectronic devices.

## Introduction

In the last decade, organic-inorganic lead-halide perovskites have revolutionized the field of optoelectronics. The unique optical and electronic properties of these materials result in high quantum, detection, and photovoltaic efficiencies, making them ideal candidates for applications as diverse as solar cells, light sources, and photodetectors [1,2,3,4,5]. All of these technologies critically depend on the details of the charge carrier dynamics and transport [1,6,7,8,9]. In the case of perovskite single crystals, photogenerated electrons and holes can diffuse for long distances, thanks to their increased tolerance against defects and to mobilities larger than those shown by thin films [10,11]. Intrinsic electric dipoles induced by the organic cations seem to facilitate the spatial separation of charges [12], while filling and accumulation of carriers at band edges influence the electron-hole recombination rate [13,14].

Another critical aspect governing the carrier dynamics is the interaction between the electronic and ionic degrees of freedom [15,16,17,18].The matrix elements of the electron-phonon coupling are in fact key in determining the intrinsic limit of carrier mobilities [19] and in defining the relaxation time of photoexcited electron-hole pairs [6]. In particular, the fact that the carrier cooling rate is surprisingly slow in lead halide perovskites has been the subject of intense debate. To explain this anomalous behavior different electron-phonon coupling mechanisms have been invoked, which are based on a form of dynamic protection that suppresses the dissipation of the electronic kinetic energy into the lattice. At low excitation densities ($n < 10^{17}$ $cm^{-3}$), it is proposed that formation of large polarons on a time scale similar to that of cooling prevents the energetic carriers from scattering with phonons and defects [20,21]. At high excitation densities ($n > 5 \times 10^{17}$ $cm^{-3}$), it is instead believed that a hot-phonon bottleneck sets in [22,23], in which charge carriers and hot optical phonons establish a dynamic equilibrium and hinder the decay of the hot phonons into modes with lower energy. To address the real-time details of the electron-phonon coupling in these materials, so far most ultrafast studies have relied on optical probes, thus revealing the impact of the ionic motion on the optical properties but providing only indirect information on the pure lattice response [16,17,21,24,25]. More recently, ultrafast structural probes have been used to explore the lattice dynamics in perovskite thin films, where the presence of localization centers (e.g., defects and grain boundaries) can significantly influence the intrinsic electron-phonon coupling [26]. In high-quality hybrid perovskite single crystals, instead a substantial lack of knowledge remains regarding the details of the electron-phonon interaction, especially on its relevant spatiotemporal scales of nanometers and few hundred femtoseconds.

Here, we set a first milestone toward this challenge by tracking the electronic and structural dynamics of a methylammonium lead iodide ($CH_3NH_3PbI_3$) single crystal. Combining ultrafast electron diffraction (UED), time-resolved surface photovoltage measurements, and time-dependent first-principles calculations allows us to provide direct signatures of a hot phonon bottleneck effect and discover a collective emission of low-energy phonons that proceeds in a regenerative manner. Our study offers an unprecedented view on the long-sought microscopic mechanism controlling the energy transfer between the excited carriers and lattice vibrations in these hybrid semiconductors.

**Structural signature of the hot-phonon bottleneck**

In our study, we use a *p*-doped single crystal of $CH_3NH_3PbI_3$ at room temperature, whose optical band gap lies around 1.47 eV [10]. Figure 1a,b shows the tetragonal unit cell together with the diffraction pattern of the (001) surface along the [110] zone-axis. To directly visualize the structural dynamics in the high excitation regime, we perform UED in the reflection geometry while pumping the crystal above the band gap (Fig. 1c). In UED, the change of diffraction intensity reflects the mean square displacement (MSD) of the atoms along the direction of the scattering vector, providing direct insights into the time evolution of the lattice response [27,28,29,30]. In a generalized Debye-Waller theory [31], the MSD scales as $\sim 1/(m\omega^2)$, where $m$ is the mass of the atoms involved in the vibration and $\omega$ is the phonon frequency. Thus, the largest diffraction intensity change is qualitatively associated with the vibration of light atoms and the excitation of low-energy phonon modes, while heavy atoms and high-energy phonons generally give a very small contribution.

In our experiments, we focus on the time evolution of the (0 0 12) Bragg reflection (marked in Fig. 1b), carefully avoiding any photoinduced sample degradation and accurately calibrating the zero-delay time at which the optical pump and electron probe pulses temporally overlap on the sample. More details are provided in the Methods section and in Supplementary Notes 1-3. Figure 1d displays the temporal evolution of the diffraction intensity upon photodoping with carrier densities of ~ $1.1\times10^{18}$ and ~ $2.2\times10^{18}$ $cm^{-3}$. We observe a time-lagged decrease of the diffraction signal, indicating that the lattice response is not triggered at zero-delay time but appears only after a lag-time, Δt, of several tens of picoseconds. Such time lag scales linearly with increasing excitation fluence, varying by a factor of two in the explored density regime (Fig. 1e). This delayed lattice response is highly anomalous if compared to the dynamics reported in other inorganic semiconductors [28,29,30,31], in which the change of the diffraction intensity is initiated at the zero-delay time and rapidly builds up in a few picoseconds (i.e. once the photocarriers have released their excess energy into high-frequency optical phonons and the latter have anharmonically decayed into lower frequency modes). It is therefore crucial to elucidate the origin of such intriguing lattice response shown by hybrid perovskite single crystals.

The absence of any change of the diffraction pattern during the lag time indicates a negligible MSD of atoms (i.e. below the sensitivity of our state-of-the-art apparatus), thus suggesting that the excitation of low-energy phonons – mediating the long-range structural response of the crystal – is initially hindered. Such a behaviour can be rationalized within a hot-phonon bottleneck scenario that describes the photophysics of hybrid perovskites under high intensity photoexcitation. Above a critical density of $5\times10^{17}$ $cm^{-3}$, as inferred from all-optical experiments [22,23], charge carriers and hot high-energy phonons reaches a dynamic equilibrium through a continuous energy exchange with no direct relaxation into low-energy modes. Blocking the phonon scattering channel toward such modes can be caused by large phononic band gaps, ferroelectric distortions, and decoupled rotations, [22] as these phenomena reduce the number of available phonon states and split phonon levels. In these conditions, when the photoexcited carrier density is high, the relaxation toward low-energy modes becomes unfavourable and the hot phonons are more likely reabsorbed by the photoexcited carriers. As a result, a dynamic freezing of the long-range lattice motion persists for tens of ps, leading to the structural lag time observed here. This interpretation is confirmed by the striking correspondence between the fluence-dependent behaviour of Δt in Fig. 1e and the progressive increase of photocarrier cooling time in a hot-phonon bottlenecked environment with increasing excitation density [22]. We note that our photoexcitation regime not only lies above the critical threshold for the occurrence of a hot-phonon bottleneck, but also exceeds the critical density for polaron dissociation in $CH_3NH_3PbI_3$ (~ $4x10^{17}cm^{-3}$) [32]. Thus, we can conclude that our results represent the structural manifestation of the hot-phonon bottleneck effect alone.

We also note that – after the lag time Δt – the UED signal exhibits a sharp decrease, which indicates the presence of a structural mechanism breaking the phonon bottleneck. As we show below, such a process is determined by the cumulative confinement of carriers at the crystal surface that eventually induces a rapid collective emission of low-energy phonons associated with the organic sub-lattice. Because the dynamics have the typical features of a regenerative emission, we denote this effect as "phonon avalanche," in analogy to the nomenclature used to describe the delayed and sudden relaxation of paramagnetic ions by the emission of phonons in a bottlenecked environment [33].

## Microscopic origin of the coupled electronic and ionic dynamics

Before demonstrating that the conditions necessary to create the phonon avalanche are met in our $CH_3NH_3PbI_3$ single crystals, we first establish the microscopic origin of the lattice change measured after $\Delta t$. In particular, we elucidate whether the signal stems from the coupling between the hot carriers and the low-energy modes, or if it results from a thermal heating process in which the photoexcited carriers have already released their excess energy before the long-range lattice response sets in. To this aim, we perform time-dependent density-functional theory (TD-DFT) calculations of the coupled electron-lattice dynamics [34]. We adopt such a method to provide direct insights into the lattice motions induced by the perturbation of the interatomic potential resulting from unrelaxed electron-hole excitations, thus mimicking the dynamics occurring at a time instant $\Delta t$. Due to the numerical complexity of the calculations, this approach cannot cover the large span of spatiotemporal scales explored during the carrier transport and lattice evolution in the hot-phonon bottleneck regime.

In the calculations, we assume an instantaneously high electronic excitation (one electron per unit cell), corresponding to the removal of an electron from the highest occupied state and its transfer to the lowest unoccupied state (further details are given in Supplementary Note 4). Figure 2a displays the spatial localization within the unit cell for the ground and the excited states: the electronic excitation mainly induces a transfer of electronic density from the I ions to the Pb ions, and in smaller extent also to the C and N ions (see Supplementary Movie 1). The analysis allows us to reveal a significant ionic displacement resulting from the strong coupling between the hot carriers and the vibrational degrees of freedom. This is indicated by the large transient atomic MSD, $\langle u^2 \rangle$, in Fig. 2b. In this quantity, we separate the contributions given by the organic ($CH_3NH_3$) and the inorganic ($PbI_3$) components and compare them to the UED data. For the highest excitation density of $2.9{\cdot}10^{18}$ $cm^{-3}$ experimentally we obtain a value of $\langle u^2 \rangle_{exp} \approx 17.2\, pm^2$. In this excitation regime (see Supplementary Note 4), the normalized MSD calculated for the organic component, $\langle u^2 \rangle_{theo}^{MA} \approx 18.85\, pm^2$ has remarkable agreement with the measured value and it is almost one order of magnitude larger than the contribution of the inorganic part, $\langle u^2 \rangle_{theo}^{PbI3} \approx 1.95\, pm^2$. Reasonably, the lighter organic unit is more mobile than the heavier inorganic cage (see also Supplementary Movie 1). Thus, the structural dynamics is mainly dominated by the low-energy vibrations of the methylammonium cations. This is consistent with an excess energy of ~ 40-80 meV that the photoexcited electrons and holes have before cooling. This value sets the upper limit to the energy of the excited phonons, which mainly include rotational and librational modes of the $CH_3NH_3$ unit, together with the vibrations of the inorganic cage [35].

Importantly, by means of molecular dynamics simulations we also calculate the MSD following thermal equilibrium trajectories, which appear to be almost an order of magnitude smaller than the values obtained above (see dashed curves in Fig. 2b and Supplementary Movie 2). Thus, we conclusively rule out a thermal lattice heating scenario behind the observed lattice response (see also Supplementary Note 5), and confirm that the modification of the interatomic potential induced by the non-equilibrium distribution of photoexcited species is the source of the complex lattice rearrangement monitored by UED via direct coupling to the low-energy phonon modes.

The fact that organic cations do play a key role in the time-delayed lattice dynamics is further confirmed by two additional considerations. The first is the analysis of the lattice response time scale. As shown in Fig. 1d, the rise time of the diffraction signal proceeds over a timescale of 1-3 ps, which is longer than our temporal resolution (~700 fs [31,36]) and consistent with the rotational motion of the $CH_3NH_3^+$ cations [35,37]. The second regards the change exhibited by the signal upon direct modification of the perovskite microstructure. In this respect, a possible strategy would rely on freezing the organic cation motion by performing low-temperature experiments. This is, however, not feasible to achieve, since lowering the sample temperature drives a structural transition into an orthorhombic unit cell, complicating the overall dynamics. An alternative approach relies on externally modifying the crystal by removing the organic cations through intense heating. We accomplish this by exposing the $CH_3NH_3PbI_3$ crystal to a train of intense infrared laser pulses for several hours, with a fluence of ~ 1.5 mJ/cm$^2$ well above the damage threshold of the organic cations. This causes significant sample degradation within the laser penetration depth, transforming the hybrid organic-inorganic crystal into a purely inorganic $PbI_X$ sample [38,39]. Figure 1f shows the resulting diffraction intensity change of the (0 0 12) reflection. We now observe a non-delayed dynamics initiating at zero-delay time with an exponentially varying phonon population, as typically seen in inorganic semiconductors where the phonon bottleneck effect is absent. Therefore, we conclude that the removal of the organic part induces a more efficient coupling between the photoexcited carriers and low-energy lattice modes via deformation potential and/or thermoelastic mechanisms.

**Carrier-driven phonon avalanche**

To understand the dynamics responsible for the structural change at $\Delta t$, we first recall that carrier-phonon scattering in inorganic semiconductors is generally described by an instantaneous logarithmic decay rate proportional to the population of excited vibrational modes. In the absence of a phonon bottleneck a rapid energy dissipation occurs, causing the phonon population to vary exponentially with time. This is the physical picture behind conventional mechanisms such as the deformation potential, thermoelastic, and inverse piezoelectric coupling [40]. However, in the case of a phonon-bottlenecked semiconductor, a different picture must be invoked. Here, the energy remains stored in the electronic system in equilibrium with the bath of higher-energy phonons until additional electron-phonon and phonon-phonon scattering channels open up, involving a larger subset of low-energy lattice modes. At this moment, the low-energy phonon population starts growing rapidly, while simultaneously increasing the instantaneous decay rate (which is proportional to the phonon occupation). Thus, the emission of such low energy modes proceeds in a regenerative manner, rapidly depleting the high-energy phonon population [33]. The net effect is a "phonon avalanche," in which the dynamics are characterized by a delayed, sudden rise of the low-energy phonon population, followed by a slower relaxation back to equilibrium.

In our experiments, we can relate the regenerative phonon emission to the cumulative confinement that the charge carriers experience at the crystal surface due to the band bending induced by *p*-doping. At the surface, progressive accumulation of photocarriers can significantly increase the local density by several orders of magnitudes with respect to the initial value determined by the photoexcitation (an approximate estimate can be obtained by considering the ratio between the optical penetration depth and the confinement region width, yielding an enhancement factor of ~ 50 in our case). The high local carrier density reached at the surface in turn leads to a modification of the interatomic potential, increasing the degree of lattice anharmonicity, enhancing dynamic disorder, and favoring more efficient phonon-phonon interactions [41,42]. This effect, together with the modified electronic and phononic band dispersions [43,44,45,46], provides a larger number of electronic and phonon states with respect to the bulk, offering more channels for the carriers to release energy into the phonon subsystem. As a consequence, when the carrier density accumulated at the surface reaches high enough values, further accumulation of energy within the electronic system is no longer favourable and the system spontaneously relaxes via the collective regenerative emission of low-energy phonons of the organic sub-lattice.

**Photocarrier transport and surface accumulation**

In the present case of *p*-doped $CH_3NH_3PbI_3$ crystals, the transport of photocarriers is governed by the band bending at the surface (see Fig. 4a). Considering the amount of *p*-doping ($N_p = 4\times10^{16}$ $cm^{-3}$) [47] and the intrinsic carrier density of our crystal ($10^{10}$ $cm^{-3}$), the surface depletion layer extends for w ~ 82.6 nm (see details in Supplementary Note 6). This results in a surface space-charge density $n_s = wN_p$ = $3.3\times10^{11}$ $cm^{-2}$, which is considerably smaller than the surface density of states ($10^{13}$-$10^{14}$ $cm^{-2}$ [48,49,50]). It implies that the Fermi level is pinned at the surface and its position is independent of doping. As a consequence, at the surface a significant downward band bending exists, which induces an average electric field $E_i$ ~ $2\times10^6$ V/m along the c-axis. Such a field acts on the photogenerated carriers as an effective driving force, inducing a drift motion along the c-axis over a distance approximately equal to the surface depletion layer and leading to the progressive accumulation of electrons at the surface and motion of holes toward the bulk (Fig. 4b).

To verify whether the dynamics of the photocarrier drift does correlate with the time lag of the diffraction response, we compare the prediction of a drift model with the fluence dependence observed for Δt. An increase in the photoexcited electron-hole density directly translates into an enhancement of carrier-carrier scattering during the electronic propagation. The mobility ($\mu$) and drift velocity ($v_d = \mu E_i$) are thus expected to decrease with increasing carrier density [51,52]. Because in this scenario the time lag would scale according to the drift time, i.e. $\Delta t \sim w/v_d$, $\Delta t$ will progressively increase with the excitation density. Consistent with this hypothesis, in our experiments we observe a monotonic dependence between the time lag and the carrier density (see Fig. 1e). Strikingly, similar quantitative agreement is also retrieved by considering the length of the depletion region $w$, the value of the intrinsic electric field $E_i$, and an electron mobility of 24 $cm^2/Vs$ [10], which yields $\Delta t$ of ~ 17.2 ps. This value is in excellent agreement with the measured 20 ps at the lowest density of $1.1\times10^{18}$ $cm^{-3}$. Despite its simplicity, the drift model captures the physical picture behind our observations. Further refinements must account for the progressive reduction of $E_i$ over time, as the band bending reduces upon surface charging, and thus the estimated final value for $\Delta t$ would be even closer to the measured one.

To obtain further evidence in favour of such a drift and confinement scenario, we also experimentally map the photovoltage dynamics of the $CH_3NH_3PbI_3$ crystal. Indeed, the photovoltage dynamics relates to the change experienced by the surface band bending upon drifting and accumulation of the photogenerated minority carriers within the surface depletion layer [53]. The signature of the photovoltage can be monitored directly using time-resolved photoelectron spectroscopy (TR-PES) [54,55] (see Fig. 3a and Methods for further details). Both TR-PES and UED investigate very similar length scales within the material and can thus be safely compared. Although the surface photovoltage affects the entire photoelectron spectrum, core-levels provide the clearest signature of such an effect. To this end, we focus on the temporal evolution of the Pb $5d_{5/2}$ shallow core level under laser excitation conditions similar to those of the UED experiment (Fig. 3b). At these fluences, the photoexcited layer (surface) density is larger than $10^{11}$ $cm^{-2}$ and thus it is comparable to the surface charge density due to the *p*-doping. Therefore, we expect a significant modification of the surface band bending. Figure 4c shows how the Pb $5d_{5/2}$ peak energy shifts as a function of pump-probe delay. We observe a positive shift of the photoelectron kinetic energy that persists for hundreds of picoseconds at both positive and negative delay times. The measured core-level shift is consistent with the downward band bending at the surface and can be ascribed to the interaction between the photoemitted electrons and the long-range dipole created outside the sample surface by the separation of optically excited electron-hole pairs. As described above, in a *p*-doped crystal the intrinsic electric field $E_i$ associated with the surface band bending produces a progressive accumulation of electrons at the surface, while holes are pushed away into the bulk. This creates a progressively accelerating dipole that increases the kinetic energy of the escaping electron, as observed experimentally. The transient modification of the surface band bending due to the motion and surface accumulation of optically excited electrons is therefore reflected into the temporal evolution recorded before time zero. Importantly, the time constant associated with such negative-delay dynamics, $\tau_{1/2}$, is related to the carrier drift time [56]. By plotting the dependence of $\tau_{1/2}$ on the excited carrier density (Fig. 4d), we observe the same trend found for Δt in the UED experiments (Fig. 1d).

## Discussion

Our results establish a direct connection between the structural time lag $\Delta t$ and the timescale of carrier drift motion at the surface of $CH_3NH_3PI_3$ single crystals (during which the hot-phonon bottleneck is active). To examine whether such drift-motion and surface-accumulation scenario is compatible with the phonon avalanche picture, we also solve a model of coupled rate equations for the carrier and phonon populations (see details in Supplementary Note 7). In the model we account for the carrier motion, phonon bottleneck, carrier accumulation, spatial confinement, and regenerative phonon emission. Figure 4c shows the solution of the model for the normalized phonon population, reproducing the salient features of our experimental observations. The model inherently contemplates the presence of a delayed response, where $\Delta t$ can be modulated by varying the transport parameters. After $\Delta t$, the phonon population shows an increase to a maximum value (avalanche), followed by a slower relaxation back to equilibrium. As detailed in Fig. 4d, this model also correctly describes the inverse correlation between $\Delta t$ and the electron mobility as expected in a drift transport picture ($\Delta t \sim w/\mu E_i$).

In conclusion, our measurements represent the first direct real-space structural signature of the hot-phonon bottleneck effect in hybrid perovskites, and unveil a hitherto unobserved effect in these materials able to overcome such a bottleneck: the phonon avalanche. This mechanism could have a significant impact on the optical and electronic properties of hybrid perovskites, likely setting an intrinsic limit to carrier mobility. Our observations, which relate the occurrence of the phonon avalanche with the accumulation and confinement of carriers at the crystal surface, can thus guide the rational design of perovskites by controlling the transport, spatial confinement, and accumulation of photoexcited charge carriers. For instance, the formation of a series of periodic scattering interfaces with separation comparable with the phonon period could favor a non-dissipative transport of charge carriers, potentially leading to higher optoelectronic efficiencies in light emitting diodes [57], gamma-ray detectors [58], and nanowire lasers [59].

**Acknowledgements**

This work was supported by the National Science Foundation and by the Air Force Office of Scientific Research in the Center for Physical Biology at Caltech funded by the Gordon and Betty Moore Foundation. J. Hu acknowledges support from Science Challenge Project (No. TZ2018001, No. TZ2016001). E.B acknowledges additional support from the Swiss National Science Foundation under fellowships P2ELP2-172290 and P400P2-183842. The perovskite crystal growth was financially supported by the Defense Threat Reduction Agency (Award No. HDTRA1-14-1-0030). M. C. acknowledges support from the European Research Council via the Advanced Grants Horizon 2020 695197 DYNAMOX. The authors gratefully acknowledge Dr. J. Tang for helpful discussions.

**Author contributions**

G. M. V., J. Hu, and A. H. Z. conceived the idea and designed research. G. M. V. and J. Hu performed the UED experiments and analyzed data. C. A. R. and M. A. performed DFT and time-dependent DFT calculations. S. P., M. P., A. C. and M. G. performed the TR-PES experiments and analyzed the data. J. Huang and H. W. grew the single crystal perovskite samples. G. M. V., J. Hu and E. B. interpreted and modelled the results together with C. A. R., M. A., A. C., M. C., and F. C.. All authors have contributed to writing and editing the paper.

**Additional information** Supplementary Information is available in the online version of the paper. Correspondence and requests for materials should be addressed to G. M. V., J. Hu, and E.B.

**Competing financial interests.** The authors declare no competing financial interests.

## METHODS

**$CH_3NH_3PbI_3$ single crystals.** The synthesis of methylammonium lead iodide single crystals, which is reported in detail in Supplementary Note 1, is composed of two phases. First, black $CH_3NH_3PbI_3$ powder is obtained by mixing a Methylamine mixture solution ($CH_3NH_2$ + HI) and a Pb mixture solution ($Pb(Ac)_{23}H_2O$ + HI) under stirring at 110 °C. Then, the powder is dissolved in γ-Butyrolactone (GBL), from which 5 mm size single crystals are grown by gradually increasing the temperature of the solution from 100 °C to 130 °C for 3-5 hours.

**Ultrafast electron diffraction (UED).** For the UED measurements, the output of a Ti:Sapphire regenerative amplifier (λ = 800 nm, 100 fs, 2 kHz) was separated in two portions: one represents the optical pump beam that is responsible for exciting the sample; the other was used to generate ultraviolet pulses by third-harmonic generation (λ = 266 nm), which irradiate a $LaB_6$ photocathode to produce ultrafast electron pulses. The diffraction patterns generated by electron scattering from the samples were recorded in far-field on a gated CCD detector, and monitored as a function of the delay time between pump and probe. For the investigation of $CH_3NH_3PbI_3$ single crystals, we adopted a grazing (0.5-2.5°) incidence geometry (reflection scheme), tracking the transient behavior of the (0 0 12) reflection. The choice of this high symmetry reflection with a large scattering vector ($s$) reflects the best compromise between maximizing the absolute diffraction intensity (which scales as $1/s$) and maximizing its relative change for a given MSD (which scales as $s^2$). Nevertheless, other reflections behaved similarly.

To minimize space charge effects, an electron pulse was designed to contain less than 300 electrons, giving a sub-ps pulse duration. In reflection geometry, the velocity mismatch between electrons and photons and the non-coaxial illumination broadened the temporal resolution. This effect was compensated by tilting the wave front of the optical pulse with respect to its propagation direction, thus preserving the sub-ps resolution. A thorough description of the experimental setup and details on the data analysis are provided in Supplementary Notes 1-2.

Particular care was taken for the calibration of the zero-delay time, i.e. the time at which the electron and laser pulses simultaneously arrived on the sample. We used multiphoton ionization from a metallic or semiconducting surface, which created a transient and localized plasma synchronous with the laser excitation able to modify the electron pulse's position and spatial profile ("plasma lensing effect"). The results obtained for the transmission and reflection geometries are reported in Supplementary Note 3. Finally, we note that within the adopted fluence range transient electric field effects on the measured diffraction pattern were negligible and did not affect the measured dynamics [60].

**Time-resolved photoelectron emission spectroscopy (TR-PES).** The surface photovoltage time-resolved traces were acquired at the Harmonium facility [61] with ASTRA solid-state end-station [54]. The ultrashort extreme ultraviolet (XUV) probe pulses were produced in the process of high harmonic generation in gas at the repetition rate of 6 kHz. A single harmonic with a photon energy of 40.3 eV was selected using a time-preserving monochromator [62]. Photoelectron spectra were collected at room temperature using a hemispherical electron energy analyzer (Specs Phoibos 150). The combined experimental energy resolution was better than 200 meV. The clean and flat sample surface required for PES measurements was obtained by mechanically cleaving the single crystals of $CH_3NH_3PbI_3$ in ultrahigh vacuum conditions. The surface was excited using the inherently synchronized fundamental infrared pulse at 1.55 eV and temporal duration of 50 fs at an incident fluence in the range 230 - 605 μJ/cm$^2$. The time delay between the XUV and infrared pulses was varied by means of a mechanical delay stage. The Pb *5d* angle-integrated signal was collected with an angular acceptance of 20° at an emission angle of 15°. The resulting spin-orbit split doublet is fitted well as a function of pump-probe delay by Gaussian functions. The fitted photoelectron peak kinetic energy is shown in Fig. 3c in a delay range between -200 ps and 200 ps around optical excitation.

**Ground-state DFT calculations.** Electronic ground state calculations were performed using a plane-wave pseudopotential approach within DFT, with an exchange-correlation functional at Perdew-Burke-Ernzerhof (PBE) level as implemented in the open-source code octopus (gitlab.com/octopus-code/octopus). For the perovskite single crystal, a cubic phase cell structure was considered. The details of the models built and the specifics of the calculations are reported in Supplementary Note 4.

**Time-dependent DFT calculations.** TD-DFT as real time Kohn-Sham propagation was employed to study the excited state properties. More details are provided in Supplementary Note 4. The initial geometries were obtained by simplifying the relaxed DFT ground state, and the core electrons were represented by Troullier-Martins pseudopotentials. The electron density and other space-dependent quantities were discretized on a three-dimensional real space grid, and the time-dependent Kohn-Sham equations were solved in real time with the finite differences method as implemented in the open-source code *octopus* (gitlab.com/octopus-code/octopus). In all the simulations, the explicit solution of the time-dependent Kohn-Sham equations was coupled to classical nuclear equations of motion via the Ehrenfest Hamiltonian. The simulations were performed in two distinct regimes. In one case the system was propagated freely from its electronic ground state, with an initially imposed random Maxwell-Boltzmann distribution of the nuclear velocities corresponding to a temperature of 300 K. These runs were meant to assess the value of the ionic MSD from their equilibrium positions as due to thermal fluctuations only. In a second set of simulations the system was initially put in an electronically excited state corresponding to a Franck-Condon transition. A sudden change in the electron density was imposed at the first time step, in order to mimic the variation of the electron population corresponding to a chosen excitation of the system. From the calculations, we followed the evolution of the total electronic density and of the ionic MSD in both cases, in order to distinguish peculiar features uniquely related to the electronic excited state and to its coupling with the ionic motions.

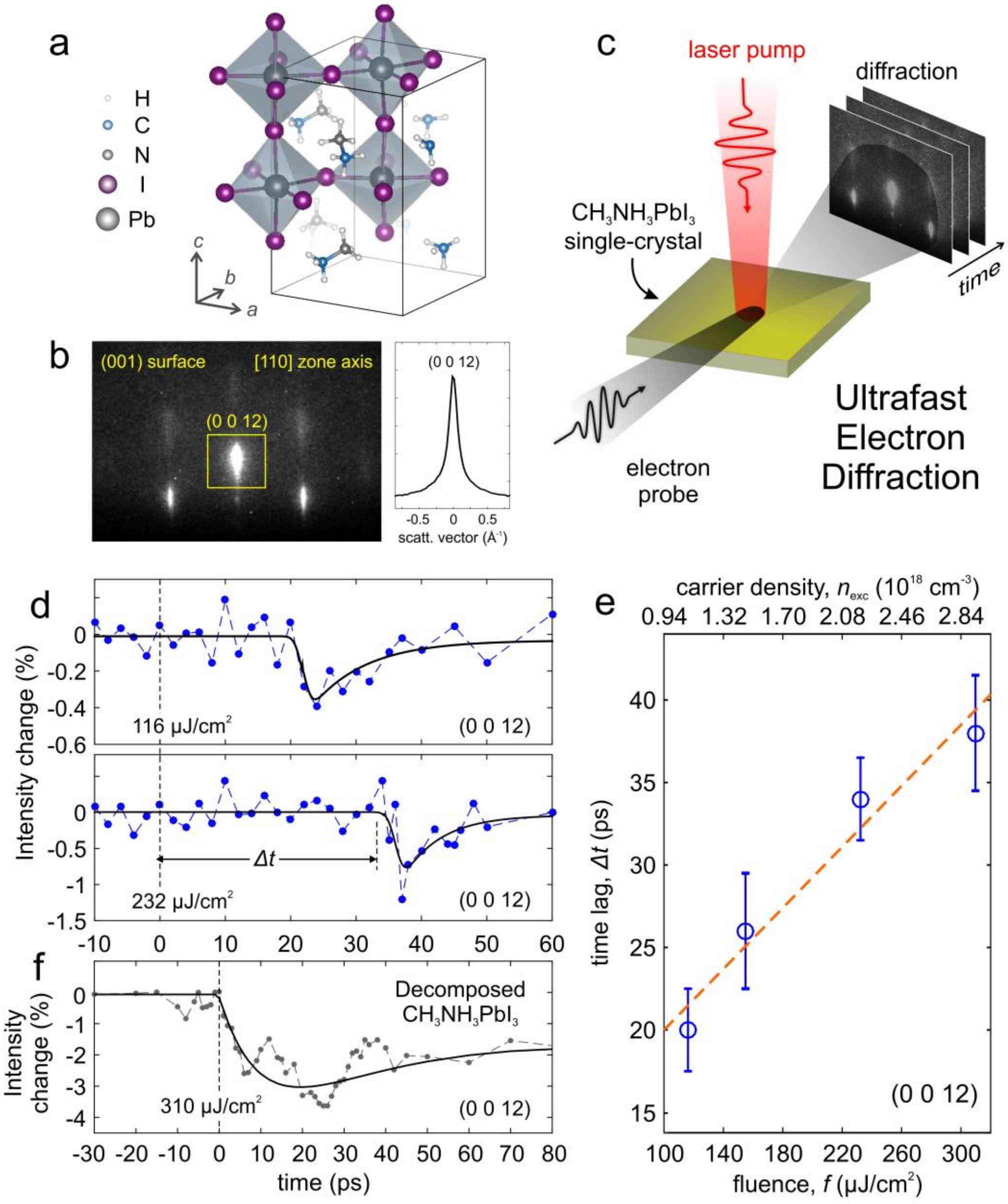

**Fig. 1. Static atomic structure and ultrafast lattice dynamics of hybrid halide perovskites.** **a**, Tetragonal unit cell of a $CH_3NH_3PbI_3$ single crystal at room temperature. **b**, Diffraction pattern of the (001) surface along the [110] zone-axis is shown at the bottom, together with the profile of the investigated (0 0 12) reflection. **c**, Schematic representation of the UED configuration in reflection geometry. **d**, Temporal evolution of the diffraction intensity for the (0 0 12) reflection upon photoexcitation with 1.55 eV light for two different incident fluences (excitation densities), 116 μJ/cm$^2$ (1.1x10$^{18}$ cm$^{-3}$) and 232 μJ/cm$^2$ (2.2x10$^{18}$ cm$^{-3}$). The zero-delay time has been calibrated using a $PbI_x$ bulk substrate (see Supplementary Note 3). **e**, The lag time, $\Delta t$, extracted from the (0 0 12) transients is plotted as a function of the excitation fluence (bottom scale) and of the corresponding excited carrier density (top scale) (the dashed line is a guide for the eye). **f**, Transient intensity change of the (0 0 12) reflection measured on a decomposed $CH_3NH_3PbI_3$ crystal as obtained by laser-annealing the perovskite crystal using a train of intense infrared laser pulses for several hours, with a fluence of ~ 1.5 mJ/cm$^2$ well above the damage threshold of the organic cations.

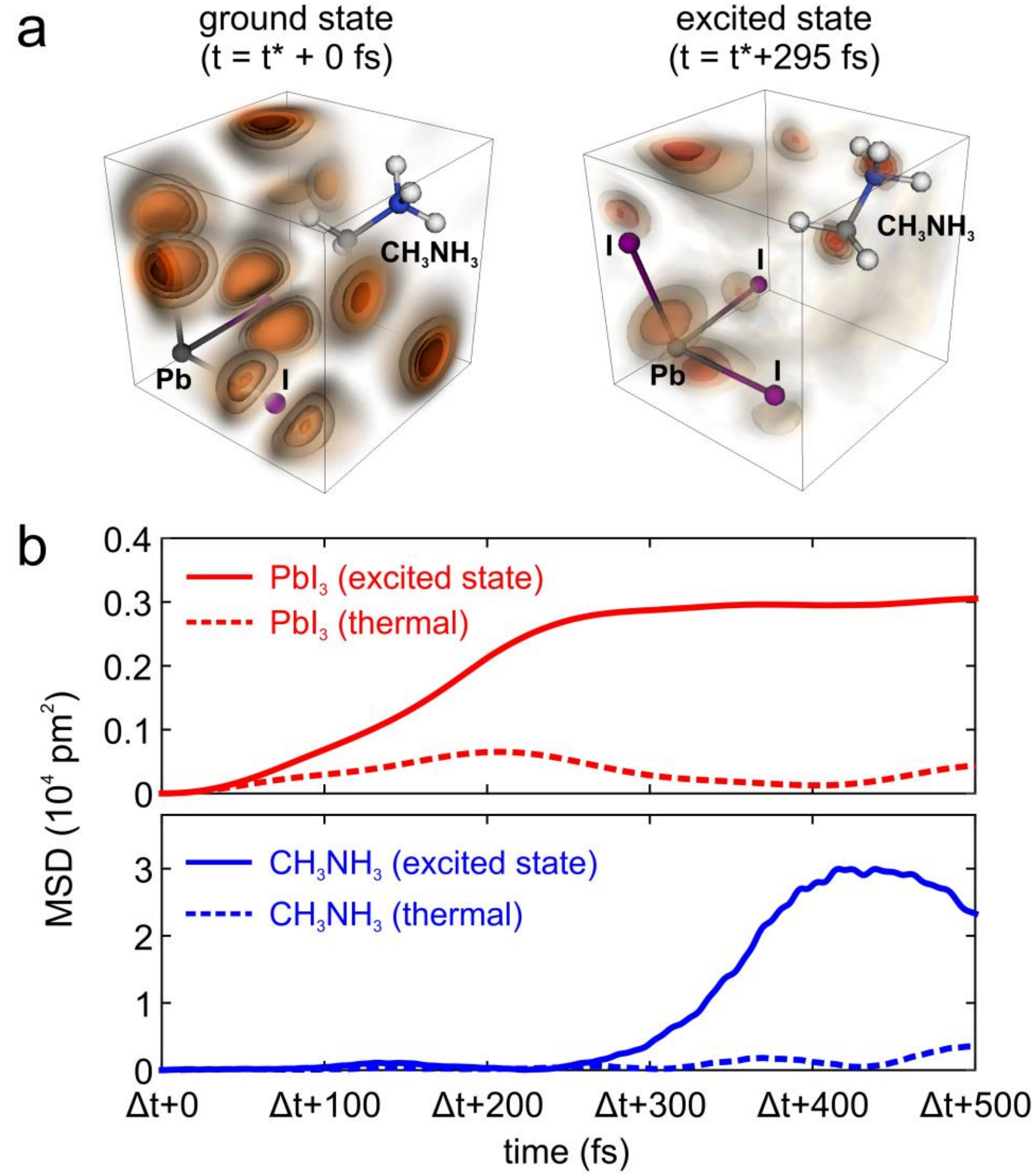

**Fig. 2. Time-dependent *ab initio* calculations of coupled electronic and nuclear dynamics.** **a**, The spatial localization of the band-edge for the ground state (left) and excited state (right) is displayed for a $CH_3NH_3PbI_3$ cubic unit cell. **b**, Calculated transient behaviour of the atomic MSD as a result of the interaction between the electronic and vibrational degrees of freedom ($\Delta t$ indicates the time of charge accumulation and spatial confinement). We distinguish the contribution to the MSD given by the organic $CH_3NH_3$ (bottom) and the inorganic $PbI_3$ (top) components. In both panels, the dashed lines define the molecular dynamics simulations following equilibrium thermal trajectories.

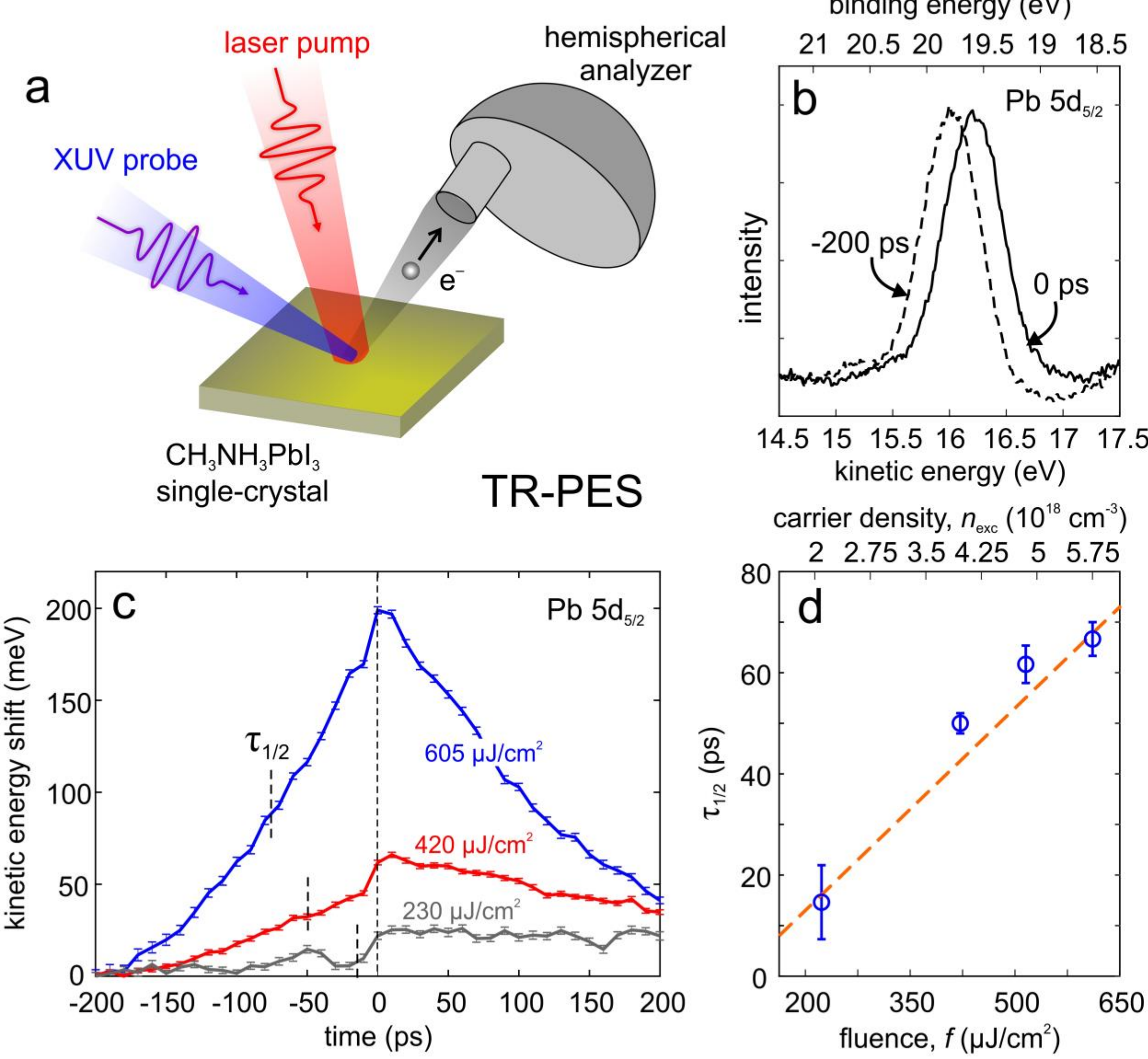

**Fig. 3. Surface photovoltage measurements in hybrid halide perovskites. a**, Schematic representation of TR-PES experiment. **b**, Lineshape of the measured Pb *5d*$_{5/2}$ shallow core level as a function of photoelectron kinetic energy (bottom scale) and core-level binding energy (top scale) for two different delay times (dashed line for t = -200 ps, and solid line for t = 0 ps). **c**, Pb *5d*$_{5/2}$ peak energy shift measured as a function of the delay time between the infrared pump and the XUV probe in the range between -200 ps and 200 ps for different excitation fluences. The vertical dashed lines correspond to the time delay at half maximum, $\tau_{1/2}$, in the negative-delay dynamics. These values are indicative of the time needed to build-up the photovoltage, and thus correlate with the carrier drift time. **d**, The observed time delay at half maximum, $\tau_{1/2}$, in the negative-delay dynamics is plotted as a function of the excitation fluence (bottom scale) and of the corresponding excited carrier density (top scale). The dashed line is a guide for the eye.

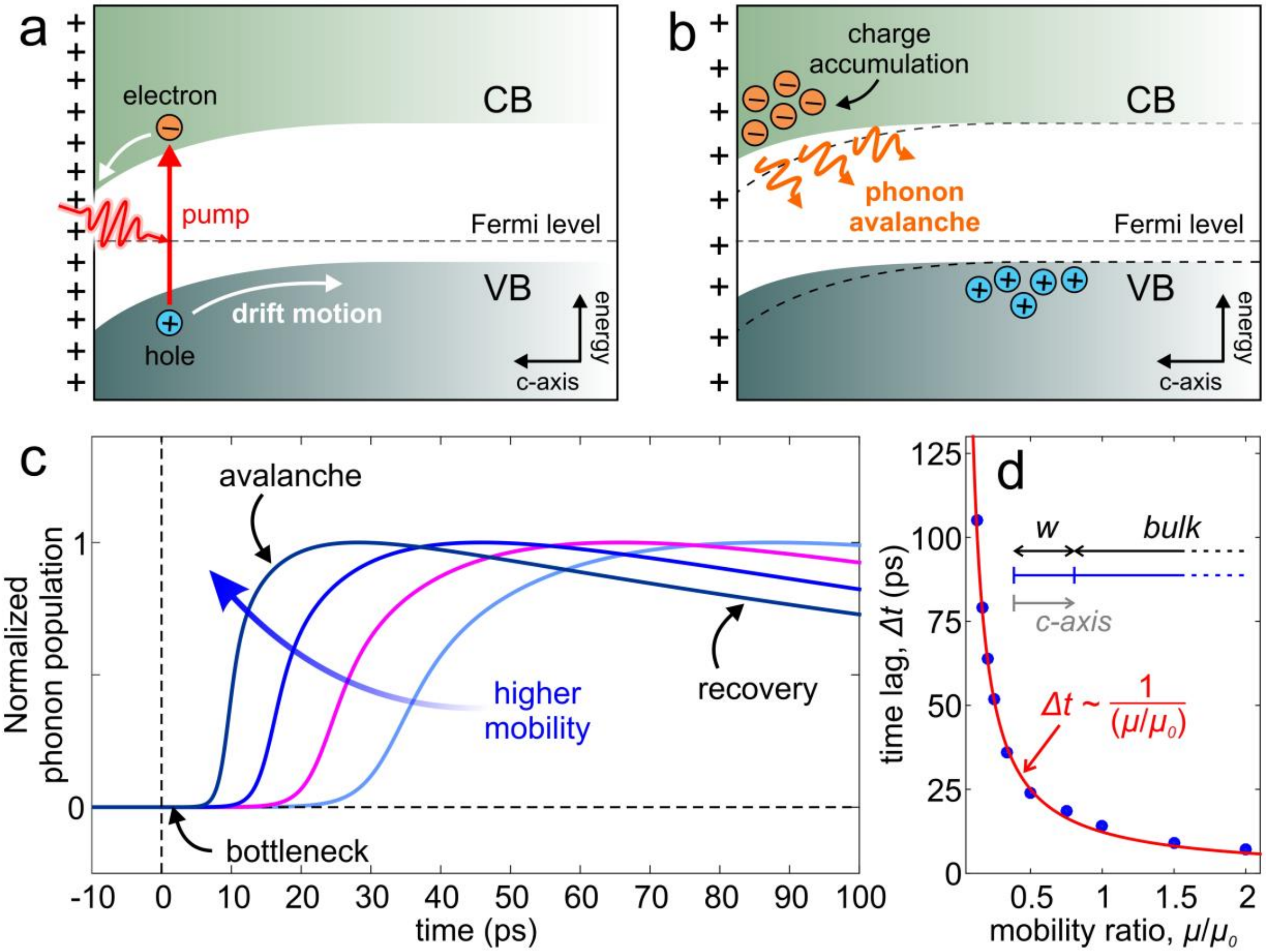

**Fig. 4. Modelling the coupled electronic-structural dynamics of lead-halide perovskites. a**, Schematic representation of the electron-hole excitation and motion in a *p*-doped hybrid perovskite single crystal. CB = conduction band, VB = valence band. **b**, As a result of the surface band bending due to the intrinsic *p*-doping, electrons accumulate at the surface, reaching local carrier densities much higher than the initial photoexcitation. At the surface, the electrons can explore a wider region of their energy landscape. The multiple decay paths now accessible to the carriers create the phonon avalanche. **c**, Simulated transient behavior of the normalized phonon population for different transport conditions using a rate-equations model including carrier drift, phonon bottleneck, carrier accumulation, spatial confinement, and regenerative phonon emission (see text and Supplementary Note 7). **d**, The numerically simulated time lag, $\Delta t$, is plotted as a function of the normalized carrier mobility, showing the dependence of the lattice response on the microscopic details of the charge motion ($\mu_0$ = 24 $cm^2$/Vs).

[20] H. Zhu, K. Miyata, Y. Fu, J. Wang, P. P. Joshi, D. Niesner, K. W. Williams, S. Jin, X.-Y. Zhu, Screening in crystalline liquids protects energetic carriers in hybrid perovskites. *Science* **353**, 1409-1413 (2016).

[21] K. Miyata, D. Meggiolaro, M. T. Trinh, P. P. Joshi, E. Mosconi, S. C. Jones, F. De Angelis, X.-Y. Zhu. Large polarons in lead halide perovskites. *Sci. Adv.* **3**, e1701217 (2017).

[22] Y. Yang, D. P. Ostrowski, R. M. France, K. Zhu, J. van de Lagemaat, J. M. Luther, M. C. Beard, Observation of a hot-phonon bottleneck in lead-iodide perovskites. *Nat. Photon.* **10**, 53-59 (2016).

[23] J. Yang, X. Wen, H. Xia, R. Sheng, Q. Ma, J. Kim, P. Tapping, T. Harada, T. W. Kee, F. Huang, Y.-B. Cheng, M. Green, A. Ho-Baillie, S. Huang, S. Shrestha, R. Patterson, G. Conibeer, Acoustic-optical phonon up-conversion and hot-phonon bottleneck in lead-halide perovskites. *Nat. Commun.* **8**, 14120 (2017).

[24] G. Batignani, G. Fumero, A. R. S. Kandada, G. Cerullo, M. Gandini, C. Ferrante, A. Petrozza, T. Scopigno, Probing femtosecond lattice displacement upon photo-carrier generation in lead halide perovskite. *Nat. Commun.* **9**, 1971 (2018).

[25] F. Thouin, D. A. Valverde-Chávez, C. Quarti, D. Cortecchia, I. Bargigia, D. Beljonne, A. Petrozza, C. Silva, A. R. S. Kandada, Phonon coherences reveal the polaronic character of excitons in two-dimensional lead halide perovskites. *Nat. Mat.* **18**, 349-356 (2019).

[26] X. Wu, L. Z. Tan, X. Shen, T. Hu, K. Miyata, M. T. Trinh, R. Li, R. Coffee, S. Liu, D. A. Egger, I. Makasyuk, Q. Zheng, A. Fry, J. S. Robinson, M. D. Smith, B. Guzelturk, H. I. Karunadasa, X. Wang, X. Zhu, L. Kronik, A. M. Rappe, A. M. Lindenberg, Light-induced picosecond rotational disordering of the inorganic sublattice in hybrid perovskites. *Sci. Adv.* **3**, 1602388 (2017).

[27] G. M. Vanacore, A. Fitzpatrick, A. H. Zewail, Four-dimensional electron microscopy: ultrafast imaging, diffraction and spectroscopy in materials science and biology. *Nano Today* **11**, 228-249 (2016).

[28] G. M. Vanacore, J. Hu, W. Liang, S. Bietti, S Sanguinetti, A. H. Zewail, Diffraction of quantum dots reveals nanoscale ultrafast energy localization. *Nano Lett.* **14**, 6148-6154 (2014).

[29] J. Hu, G. M. Vanacore, A. Cepellotti, N. Marzari, A. H. Zewail, Rippling ultrafast dynamics of suspended 2D monolayers, graphene. *Proc. Natl. Acad. Sci. USA* **113**, E6555-E6561 (2016).

[30] G. M. Vanacore, R. M. van der Veen, A. H. Zewail, Origin of axial and radial expansions in carbon nanotubes revealed by ultrafast diffraction and spectroscopy. *ACS Nano* **9**, 1721–1729 (2015).

[31] S. Schäfer, W. Liang, A. H. Zewail, Primary structural dynamics in graphite. *New J. Phys.* **13**, 063030 (2011).

[32] J. M. Frost, L. D. Whalley, A. Walsh, Slow cooling of hot polarons in halide perovskite solar cells. *ACS Energy Lett.* **2**, 2647−2652 (2017).

[33] W. J. Brya, P. Wagner, Dynamic interaction between paramagnetic ions and resonant phonons in a bottlenecked lattice. *Phys. Rev.* **157** 400-410 (1967).

[34] S. M. Falke, C. A. Rozzi, D. Brida, M. Maiuri, M. Amato, E. Sommer, A. De Sio, A. Rubio, G. Cerullo, E. Molinari, C. Lienau, Coherent ultrafast charge transfer in an organic photovoltaic blend. *Science* **344**, 1001-1005 (2014).

[35] F. Brivio, J. M. Frost, J. M. Skelton, A. J. Jackson, O. J. Weber, M. T. Weller, A. R. Goni, A. M. A. Leguy, P. R. F. Barnes, A. Walsh, Lattice dynamics and vibrational spectra of the orthorhombic, tetragonal, and cubic phases of methylammonium lead iodide. *Phys. Rev. B* **92**, 144308 (2015).

[36] F. Pennacchio, G. M. Vanacore, G. F. Mancini, M. Oppermann, R. Jayaraman, P. Musumeci, P. Baum, F. Carbone. Design and implementation of an optimal laser pulse front tilting scheme for ultrafast electron diffraction in reflection geometry with high temporal resolution. *Struct. Dyn.* **4**, 044032 (2017).

[37] A. A. Bakulin, O. Selig, H. J. Bakker, Y. L.A. Rezus, C. Müller, T. Glaser, R. Lovrincic, Z. Sun, Z. Chen, A. Walsh, J. M. Frost, T. L. C. Jansen, Real-time observation of organic cation reorientation in methylammonium lead iodide perovskites. *J. Phys. Chem. Lett.* **6**, 3663-3669 (2015).

[57] H. Cho, et al. Overcoming the electroluminescence efficiency limitations of perovskite light-emitting diodes, *Science* **350**, 1222-1225 (2015).

[58] S. Yakunin, et al. Detection of gamma photons using solution-grown single crystals of hybrid lead halide perovskites. *Nat. Photon.* **10**, 585-589 (2016).

[59] H. Zhu, et al. Lead halide perovskite nanowire lasers with low lasing thresholds and high quality factors. *Nat. Mater.* **14**, 636-642 (2015).

[60] S. Schäfer, W. Liang, A. H. Zewail, Structural dynamics and transient electric-field effects in ultrafast electron diffraction from surfaces. *Chem. Phys. Lett.* **493**, 11-18 (2010).

[61] C. A. Arrell et al., Harmonium: An ultrafast vacuum ultraviolet facility, *CHIMIA* **71**, 268-272, (2017).

[62] J. Ojeda et al., Harmonium: A pulse preserving source of monochromatic extreme ultraviolet (30–110 eV) radiation for ultrafast photoelectron spectroscopy of liquids. *Struc. Dyn.* **3**, 023602 (2015).

# Supplementary Information for:

# Observation of a phonon avalanche in highly photoexcited hybrid perovskite single crystals

## S1. Sample preparation

*$CH_3NH_3PbI_3$ powder preparation*: 2.5 g of $Pb(Ac)_{23}H_2O$ (99%, Alfa Aesar) was dissolved into 7.6 ml HI (57% w/w aq. sol., stabilized with 1.5% hypophosphorous acid, Alfa Aesar) at 110 °C under stirring; 6.6 mM Methylamine ($CH_3NH_2$) (40% w/w aq. soln., Alfa Aesar) was dissolved into 0.87 ml HI under stirring for 10 minutes. The Methylamine mixture solution was then gradually dripped into the Pb mixture solution under stirring at 110 °C, and black $CH_3NH_3PbI_3$ powder was obtained at the bottom of the solution. The solution was kept stirring at 80 °C for 1 hour, then the black powder was separated from the solution and washed with diethyl ether for at least three times. The $CH_3NH_3PbI_3$ powder was then dried under vacuum for overnight.

*$CH_3NH_3PbI_3$ single crystal synthesis*: 3.1 g $CH_3NH_3PbI_3$ powder was dissolved into 5 ml γ-Butyrolactone (GBL), and the solution was filtered by 0.45 µm filter. 5 mm size $CH_3NH_3PbI_3$ single crystal was grown by gradually increasing the temperature of the solution from 100 °C to 130 °C for 3~5 hours.

## S2. Experiment and methods

### *S2.1 UED experiment*

The setup for UED experiments at Caltech has already been described in details elsewhere [1,2] and is schematically shown in Fig. S1. Briefly, a laser-pump/electron-probe scheme is adopted, working in stroboscopic mode with a variable delay time between pump and probe. Ultrafast electron pulses are generated in a photoelectron gun (Kimball Physics Inc.) after irradiation of a $LaB_6$ photocathode with femtosecond UV pulses (λ = 266 nm) generated by frequency-tripling the output of a Ti:sapphire regenerative amplifier (λ = 800 nm, 100 fs, 2 kHz).

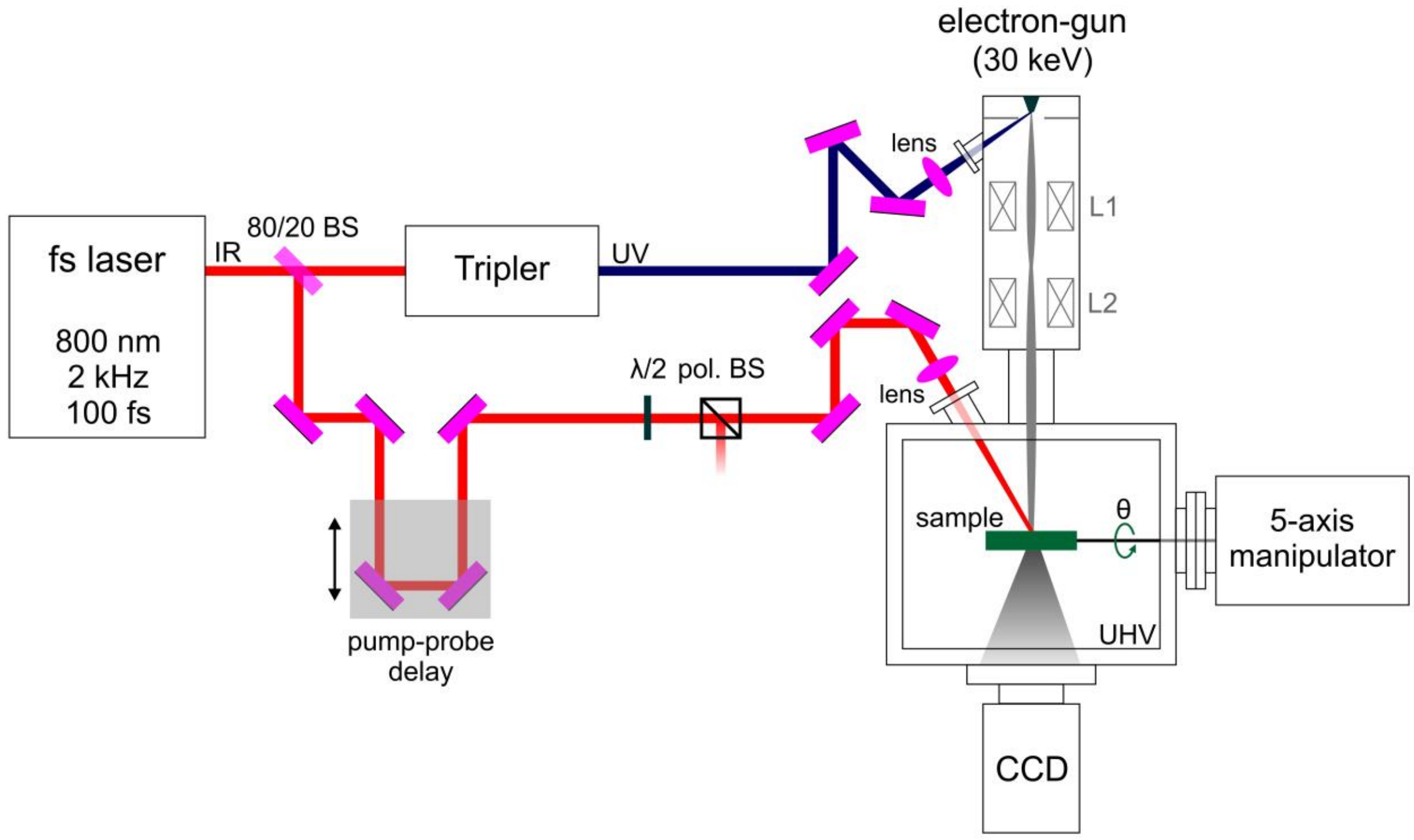

**Fig. S1. Schematic representation of the experimental setup for ultrafast electron diffraction at Caltech.** Ultrashort electron pulses are created by the UV beam at λ = 266 nm, and are used to probe the dynamics of the sample excited by the optical pump at λ = 800 nm. (IR: infrared; UV: ultraviolet; BS: beam splitter; L1 and L2: condenser and objective electrostatic lenses, respectively; CCD: charged coupled device; UHV: ultra-high vacuum).

For the investigation of $CH_3NH_3PbI_3$ single crystals we adopt a grazing (0.5-2.5°) incidence geometry (reflection scheme) with electron energy of 30 keV. The electron beam has a transverse spot size of ~ 100 µm at the sample position. The dynamics are initiated by femtosecond laser pulses at λ = 800 nm with incident fluence varying between 116 and 310 µJ/cm$^2$ arriving on the sample surface in normal incidence. The fluence of the excitation laser beam at the sample position was calibrated by scanning a knife edge across the laser profile (~ 1 mm transverse spot size) and recording the passing residual pulse energy. In order to minimize space charge effects, an electron pulse contains less than 300 electrons, giving a sub-picosecond pulse duration [3]. In the reflection scheme, the velocity mismatch and the non-coaxial geometry between electrons and photons are responsible for a different arrival time of the electrons at different regions of the sample surface illuminated by the pump pulse. This effect is compensated for by tilting the wavefront of the optical pulse with respect to its propagation direction [4], thus preserving the sub-ps temporal resolution. The relative delay time between laser-pump and electron-probe is controlled by a linear delay stage inserted into the optical path of the pump beam.

The electron diffraction patterns generated by electron scattering from the samples are recorded on a gated MCP/phosphor-screen/CCD assembly, and monitored in stroboscopic mode as a function of the delay time between pump and probe. To obtain a good signal-to-noise ratio, more than 100 temporal scans have been acquired over different sample positions, and in each scan every diffraction pattern was averaged over 20,000-40,000 reproducible events. The large spot-size of the optical pump (~ 1 mm) and the low repetition rate (2 kHz) allowed to reduce radiation damage caused by cumulative and localized heating. Because of the low fluences used, transient electric field (surface charging) effect [5] was found to be insignificant for all measurements.

*S2.2 Diffraction analysis*

We monitored the diffraction pattern obtained from the (001) surface along the [110] zone-axis (see Fig. 1). This is composed of a series of diffraction spots assigned to the $\{00\ell\}$ and $\{11\ell\}$ reflections of the tetragonal lattice. To quantitatively study the dynamics, we selected a specific region of interest within the diffraction pattern around a certain diffraction spot (for instance, (0 0 12)), from which a one-dimensional peak profile was obtained. The peak was fitted using a linear background and a Lorentzian function, from which we could retrieve the diffraction intensity and peak position of the selected reflection.

## S3. Time-zero determination

Particular care has been taken for the calibration of the zero-delay time, i.e. the time at which the electron pulse and the laser pulse simultaneously arrive on the sample. We used multiphoton ionization from a metallic or semiconducting surface, which is responsible for the so-called "plasma lensing effect". When the surface is subjected to an intense femtosecond laser pulse, photoelectrons are generated to form a transient and localized plasma synchronous with the laser excitation. This plasma is able to act as convergent or divergent lens, changing the electron pulse's position and spatial profile [4]. Using this method the zero-time is obtained with an accuracy of about 1 ps [6].

To increase the reliability of the time-zero determination we performed the calibration both in a transmission geometry and in a reflection geometry. For experiments in transmission geometry, we used a 2-3 nm Au thin film on a 300-mesh gold TEM grid (purchased from Ted Pella Inc.). Here, the electron beam arrives on the sample in normal incidence, while the laser impinges with a 45° incidence angle. For the calibration in reflection geometry, we have subjected the $CH_3NH_3PbI_3$ single crystal to a long-time illumination in vacuum with the 800 nm beam at a fluence of 1.56 mJ/cm$^2$. This results in a significant degradation of the perovskite crystal within the laser penetration depth, which transforms it into a purely inorganic $PbI_x$ crystal [7], able to generate an intense plasma. In this case, the electron beam is set to be tangential to the sample surface, while the optical pulse with a tilted wavefront arrives in normal incidence.

Figs. S2a shows the typical transient behavior of the direct electron beam deflection for the transmission scheme under a laser excitation at $\lambda_p$ = 800 nm (4.6 mJ/cm$^2$). In Fig. S2b, the transient deflection measured for the reflection geometry is reported for a laser excitation at $\lambda_p$ = 800 nm (1.56 mJ/cm$^2$). For an accurate estimation, we evaluated the zero-time from all transients by linearly fitting the initial slope and determining the crossing with the base line. This calibration has been repeated for every set of experiments, and the estimated values have been used as reference for the measured transients.

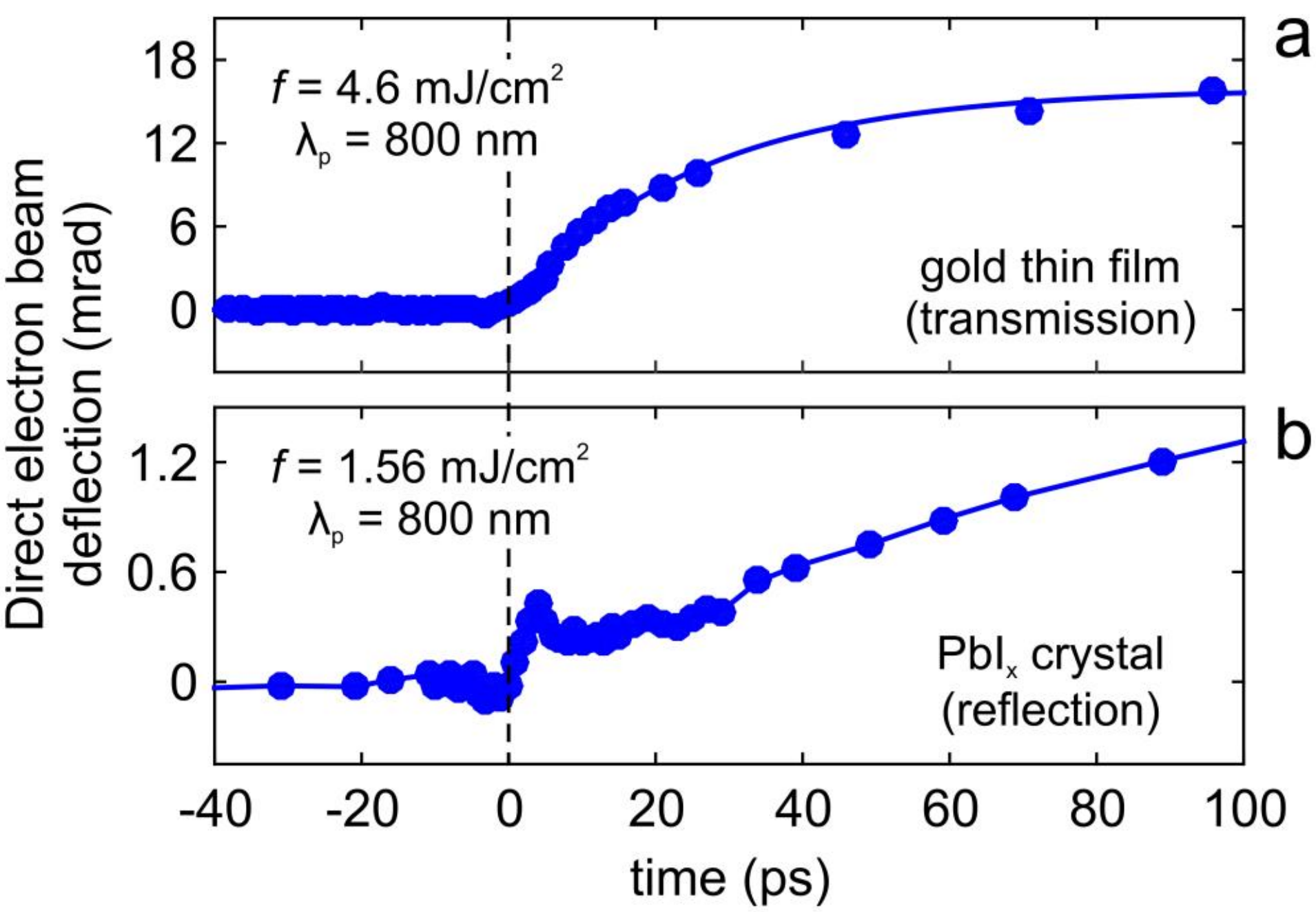

**Fig. S2. Zero-time calibration.** (**a**) Transient behavior of the direct electron beam deflection when transmitted through a 2-3 nm thick gold thin film on a gold TEM grid. The laser excitation is 4.6 mJ/cm$^2$ at $\lambda_p$ = 800 nm. (**b**) Transient deflection measured for the reflection geometry when the direct electron beam is tangential to the surface of a $PbI_x$ bulk crystal. The laser excitation is 1.56 mJ/cm$^2$ at $\lambda_p$ = 800 nm. For an accurate estimation, the zero-time from each transient is estimated by linearly fitting the initial slope and determining the crossing with the base line.

## S4. Time-dependent DFT calculations

*S4.1 Stationary states*

Ground state calculations were performed using a plane-wave pseudopotential approach within Density Functional Theory (DFT) as implemented in the open-source code (gitlab.com/octopus-code/octopus) [8,9,10]. The perovskite cubic phase cell structure with [111] alignment of the methylammonium group was taken from [11], where it was obtained by geometry relaxation at the PBE.

*S4.2 Time-dependent simulations*

Time Dependent Density Functional Theory (TDDFT) as real time Kohn-Sham propagation was employed to study the excited state properties of methylammonium lead iodide perovskite ($CH_3NH_3PbI_3$). The initial geometry was obtained by simplifying the DFT ground state relaxed geometries described in the previous section. The core electrons were represented by means of Troullier-Martins pseudopotentials. The electron density and the other space-dependent quantities were discretized on a three-dimensional real space grid with a spacing of 0.19 Å, and the time-dependent Kohn-Sham equations were solved in real time with the finite differences method as implemented in the open-source code octopus (gitlab.com/octopus-code/octopus) [9,12].

In all simulations the explicit solution of the time-dependent Kohn-Sham equations was coupled to classical nuclear equations of motion via the Ehrenfest Hamiltonian, which provides a quantum force term as a mean field back reaction of the electrons onto the nuclei. In particular the formalism described in Ref. [13] was adopted.

The time-dependent simulations were performed in two distinct regimes. In one case the system was propagated freely from its electronic ground state, with an initially imposed random Maxwell-Boltzmann distribution of the nuclear velocities corresponding to a temperature of 300 K. These runs were mostly meant to assess the value of the nuclear mean square displacements (MSD) from their equilibrium positions as due to thermal fluctuations only. In a second set of simulations the system was initially put in an electronically excited state corresponding to a Franck-Condon transition. A sudden change in the electron density was imposed at the first time step in order to mimic the variation of the electron population corresponding to a chosen excitation of the system. For consistency, the initial nuclear random velocities were kept as in the thermal equilibrium condition. Monitoring the system evolution and comparing the two runs, we were able to distinguish peculiar features that are uniquely related to the electronic excited state, and to its coupling with the nuclear motions.

In particular, we followed the evolution of the total electronic density (and of its projections on the time-dependent Kohn-Sham orbitals), and of some other relevant geometrical parameters. The nuclear MSD were calculated in groups to highlight significant differences between the behaviors of subunits of the systems.

For the calculations we kept the original perovskite unit formula, but we reduced the unit cell to cubic symmetry, with a side of 6.347 Å. The methylammonium molecule was initially oriented along the [111] direction, and periodic boundary conditions were applied on a 6x6x6 k-point grid. The time step for the simulation was set to match the electronic dynamics time scale at 1.3 as, and the time propagation was extended up to 600 fs by considering the Γ point only. The TDDFT functional of choice in this case was Perdew-Burke-Ernzerhof (PBE). The adiabatic approximation for the TDDFT kernel was maintained.

At the first time step, an approximate electronic excited state was built by removing one electron from the valence band (mostly localized on the Iodine atoms) and promoting it to the conduction band (mostly localized on Lead atom). MSD were calculated for two groups of atoms, namely for the $PbI_3$ cage and for the methylammonium molecule. Although the latter maintains an overall higher mobility in all runs, a striking difference is observed between the thermal and the photoexcited trajectories. In particular, a strong orthorhombic distortion of the Pb-I cage is observed, together with a wide libration of the methylammonium molecule. As a result the dominant direction of the *p* state axis changes, mostly correlating to the change in position of the $NH_3$ cation. See Fig. 2 in the main article, and comments therein.

**Supplementary Movie M1. Excited state dynamics of perovskite.**
The excited state electron density of a cubic cell methylammonium-lead-triiodide perovskite is projected into the conduction band edge orbital. An electron has been promoted from the valence band edge, mainly localized on the I atoms into the lowest energy conduction band state, mostly having Pb character at time zero, and the following time evolution is displayed. The orientation of the excited state is correlated with the overall cell polarization, which, in turn, is governed by huge changes both in the Pb-I cage distortion and the organic molecule libration.

**Supplementary Movie M2. Thermal dynamics of perovskite**
This movie clip shows the time propagation of the valence band edge projected charge density. During this evolution the system remains in its electronic ground state, which adiabatically follows the nuclear configuration in the ground state potential energy surface. Thermal fluctuations induce nuclear movement and distortions overall much less prominent than the ones observed in the excited state.

*S4.3 Quantitative comparison between experiment and calculations*

Because the change of diffraction intensity can be directly associated to the mean-square displacement (MSD), $\langle u^2 \rangle$, along the direction of the scattering vector, a quantitative comparison between experiments and calculations can be retrieved. It is, however, worth noting that in the calculations we adopt an excitation of one electron per cell, and thus the theoretical MSD have to be properly normalized to the effective charge excitation per unit cell at the effective experimental photoexcitation. For the largest incident fluence used (310 *μJ/cm*$^2$), the excited number of electrons per unit cell is 6.5x10$^{-4}$. The normalized values of the calculated MSDs for the methylammonium, $CH_3NH_3$, and for the lead iodide, $PbI_3$, are thus: $\langle u^2 \rangle_{theo}^{CH_3NH_3} \approx 2.9 \cdot 10^4\ \mathrm{pm}^2 \cdot 6.5 \cdot 10^{-4} = 18.85\ \mathrm{pm}^2$ and $\langle u^2 \rangle_{theo}^{PbI_3} \approx 0.3 \cdot 10^4\ \mathrm{pm}^2 \cdot 6.5 \cdot 10^{-4} = 1.95\ \mathrm{pm}^2$, respectively. Considering an intensity change of $\sim 3\%$ for the (0 0 12) reflection at 310 *μJ/cm*$^2$, the experimentally observed MSD is $\langle u^2 \rangle_{theo}^{CH_3NH_3PbI_3} \approx 17.2\ \mathrm{pm}^2$, which is close to the value calculated for the highly mobile organic component.

## S5. Equilibrium heating model

According to the equilibrium heating model [14,15] the temperature increase, $\Delta T$, induced by the laser irradiation can be calculated from:

$$n_{exc}(h\upsilon - E_g) = \int_{T_0}^{T_0+\Delta T} \rho c_p(T) dT \qquad (1)$$

where $n_{exc}$ is the photoexcited carrier density, $T_0 = 300\ \mathrm{K}$ is the initial temperature, $E_g$ is the band gap (absorption onset), $\rho$ is the mass density, and $c_p(T)$ is the temperature-dependent specific heat. The photoexcited carrier density depends on the incident fluence, $f$, through the equation:

$$n_{exc} = \frac{f\alpha(1-R)}{h\upsilon} \qquad (2)$$

where $h\upsilon$ is the photon energy, $\alpha$ is the absorption coefficient, and $R$ is the reflection coefficient. At low fluence, we have that $\Delta T \ll T_0$ and therefore $c_p(T) \sim c_p(T_0)$, leading to:

$$\Delta T \approx \frac{n_{exc}\left(h\nu - E_g\right)}{\rho c_p(T_0)} \tag{3}$$

The thermal motions for a given temperature increase, $\Delta T$, induce a loss of interference that results in a reduction of the diffraction intensity, quantitatively described by the Debye-Waller relation:

$$\ln\left(\frac{I_0}{I}\right) = 2[W(T,\Theta_D) - W(T_0,\Theta_D)] \tag{4}$$

where $T = T_0 + \Delta T$, $\Theta_D$ is the Debye temperature, $W$ is the Debye-Waller factor. The Debye-Waller factor can be expressed as:

$$W(T,\Theta_D) = \frac{3\hbar^2 s^2}{2MK_B\Theta_D}\left[\frac{1}{4} + \left(\frac{T}{\Theta_D}\right)^2 \int_0^{\Theta_D/T} \frac{\psi}{\exp(\psi) - 1} d\psi\right] \tag{5}$$

where $M$ is the average atomic mass of the atoms within the unit cell, $K_B$ is the Boltzmann constant, and $\hbar$ is the Planck constant. Using the parameters reported in the table S1 below and in hypothesis that all the incident optical pump power is converted into heating the system, we calculate a temperature jump $\Delta T$ of 0.03 K for a $CH_3NH_3PbI_3$ single crystal at the highest fluence used. This would induce an intensity change $I/I_0$ of 0.08 % for the (0 0 12) reflection, which is almost one order of magnitude smaller than the measured value.

**Table S1**

| **Material property** | **$CH_3NH_3PbI_3$ (single crystal)** |
|---|---|
| mass density, $\rho$ | $4.127 \cdot 10^3$ kg/m$^3$ |
| photon energy, $h\nu$ | 1.55 eV |
| absorption coefficient, $\alpha$ (at $h\nu$) | $2.6 \cdot 10^5$ m$^{-1}$ [17] |
| reflection coefficient, $R$ | 0.1 [16] |
| band gap (absorption onset), $E_g$ | 1.47 eV [17] |
| specific heat, $c_p$ (at 300 K) | 302.88 J·kg$^{-1}$·K$^{-1}$ [18,19] |
| Debye temperature, $\Theta_D$ | 120 K [18,20] |

## S6. Band bending in p-doped $CH_3NH_3PbI_3$ single crystal

In this section we will estimate the band bending for a p-doped MAPbI3 single crystal. Considering a p-type doping $N_p$ of 4x10$^{16}$ cm$^{-3}$ as obtained with our growth method [21] and a typical intrinsic carrier density $n_i$ of 10$^{10}$ cm$^{-3}$, the Fermi level shift is given by:

$$\Delta E = \left| E_F^{(p)} - E_F^{(i)} \right| = K_B T ln\left(\frac{N_p}{n_i}\right) = 0.38\, eV \quad (6)$$

The surface depletion layer, $w$, can be thus calculated as:

$$w = \sqrt{\frac{2\varepsilon\varepsilon_0 \Delta E}{e^2 N_p}} = 82.6\, nm \quad (7)$$

where we have used $\varepsilon = 6.5$ [22] and $\varepsilon_0$ is the vacuum permittivity. This results in a surface charge density $n_S = wN_p$ = 3.3x10$^{15}$ m$^{-2}$ that is considerably smaller than the surface density of states of about 10$^{17}$-10$^{18}$ m$^{-2}$ [23,24,25], implying that the Fermi level is pinned at the surface and its position is independent on the doping. Such situation induces a significant downward bending of the bands at the crystal surface producing an intrinsic electric field, $\mathcal{E}_i$, along the c-axis that varies linearly with the distance from the surface and reaches zero at a depth equal to w. The average value of such field is thus: $\langle \mathcal{E}_i \rangle = \mathcal{E}_i^{max}/2 \approx \Delta E/2ew$, which is on the order of 2x10$^6$ V/m. Such field would thus act on the photo-created electrons and holes as an effective driving force inducing a drift motion along the c-axis over a distance approximately equal to the surface depletion layer (w): electrons will progressively accumulate at the surface, while holes will move toward the bulk.

## S7. Modelling phonon-bottleneck and phonon-avalanche

The simplest model able to describe the observed phonon dynamics is based on coupled rate equations for the carrier and phonon populations. It is a modified version of the one developed in Ref. [26] used to describe the regenerative phonon emission in a bottlenecked lattice, where we also include the carrier diffusion within the materials. The equations are written in terms of the normalized phonon population, *p*, and the differential carrier population $\tilde{n} = n_{exc} - n_{GS}$, where $n_{exc}$ is the excited carrier density and $n_{GS}$ is the ground state carrier population.

$$\frac{\partial \tilde{n}(x,t)}{\partial t} = Q(x,t) + v_d \frac{\partial \tilde{n}(x,t)}{\partial x} - A \frac{\tilde{n}(p(x,t)+1)}{\tau_D} L(x)$$
$$\frac{\partial p(x,t)}{\partial t} = \frac{A}{\rho_{ph}} \frac{\tilde{n}(p(x,t)+1)}{\tau_D} L(x) - \frac{p(x,t)}{\tau_p} \tag{6}$$

where the source term $Q(t)$ is written as:

$$Q(x,t) = \frac{f\alpha(1-R)}{h\nu} \frac{\mathrm{sech}^2(t/\tau_{Las})}{2\tau_{Las}} \tag{7}$$

and $\tau_{Las} = 100$ fs is the laser pulse width. In Eq. (6), $v_d = \mu\mathcal{E}_i$ is the drift velocity, where $\mu$ is the electron mobility and $\mathcal{E}_i$ is the intrinsic electric field due to the surface band bending acting over a length scale given by the surface depletion layer, $w$. The parameter $A$ is the bottleneck factor and modulates the decay of the charge carriers during their diffusion, $\rho_{ph}$ is the available phonon density, and $\tau_p$ is the phonon dissipation time constant and describe the rate at which the lattice goes back to the equilibrium.

The function $L(x)$ describes the carrier localization at the surface or interface. Here the carriers can accumulate and reach densities much larger than the initial photoexcitation. In this condition they would explore a larger region of their energy landscape and multiple decay channels can open up to the phonon subsystem. For the hybrid perovskite single-crystal studied here, the growth method induces an intrinsic *p*-doping, and thus defects can trap the majority carriers close to the surface and induce the formation of a depletion layer of size $w \sim 82.6$ nm, which creates a significant downward band bending at the surface. Because of the presence of such band bending an intrinsic electric field, $\mathcal{E}_i$, is formed over the depletion layer $w$, and thus photoexcited electrons can preferentially accumulates at the crystal surface while hole are pushed away into the bulk. In this case, $L(x)$ can be given in first approximation by:

$$L(x) = H[x] - H[x - w] \quad (8)$$

where $H[\dots]$ is the Heaviside step function.

To simulate the effect on the lattice dynamics by the modulation of the kinetic conditions of carrier transport, we have calculated the transient response of the system for different values of the drift velocity $v_d$. For $CH_3NH_3PbI_3$ single-crystals the carrier mobility $\mu$, and therefore the drift velocity, exhibit an inverse correlation on the excited carrier density [27]. This results from an enhanced carrier-carrier scattering at large excitation fluences, and thus induces a significant modulation of the diffusion time constant.

With initial conditions given by $\tilde{n}(x,0) = 0$ and $p(x,0) = 0$ and boundary conditions defined by the surface, we numerically solved the model described above for the specified domain geometry and calculated the temporal behavior of the carrier and phonon populations. In Fig. 4 of the main text, we report the normalized phonon population averaged at the surface or interface for different values of the drift velocity. After an initial lag time determined by the transport parameters where the phonon bottleneck dominates, a sharp increase of the phonon population is observed when carriers accumulates at the surface or interface, which describes the regenerative emission of phonons (avalanche). Finally, the system relaxes back toward the equilibrium on a longer timescale.